\documentclass[
superscriptaddress,
reprint,
amsmath,amssymb,
aps,
prb,
]{revtex4-2}

\usepackage{graphicx}
\usepackage{dcolumn}
\usepackage{bm}
\usepackage{hyperref}

\begin{document}

\title{First-principle based Floquet engineering of solids in the velocity gauge}

\author{Vishal Tiwari}
\affiliation{Department of Chemistry, University of Rochester, Rochester, New York 14627, USA}

\author{Ignacio Franco}
\affiliation{Department of Chemistry, University of Rochester, Rochester, New York 14627, USA}
\affiliation{Department of Physics, University of Rochester, Rochester, New York 14627, USA}

\email{ignacio.franco@rochester.edu}

\date{\today} 

\pagebreak

\begin{abstract}

We introduce a practical and accurate strategy to capture light-matter interactions using the Floquet formalism in the velocity gauge in combination with realistic first-principle models of solids. The velocity gauge, defined by the linear coupling to the vector potential, is a standard method to capture the light-matter interaction in solids. However, its use with first-principle models has been limited by the challenging fact that it requires a large number of bands for convergence and its incompatibility with non-local pseudopotential plane wave methods. To improve its convergence properties, we explicitly take into account the truncation of Hilbert space in the construction of the Floquet Hamiltonian in the velocity gauge. To avoid the incompatibility with the pseudopotentials, we base our computations on generalized tight-binding Hamiltonians derived from first-principles through maximally-localized Wannier functions. We exemplify the approach by computing the optical absorption spectra of laser-dressed \textit{trans}-polyacetylene chain using realistic electronic structure. We show that, by proceeding in this way, Floquet consideration involving the truncated Hilbert spaces reproduces the full basis calculations with only a few bands and with significantly reduced computation time. The strategy has been implemented in \textsc{FloqticS}, a general code for the Floquet engineering of the optical properties of materials. Overall, this work introduces a useful theoretical tool to realize Floquet engineering of realistic solids in the velocity gauge.

\end{abstract}

\maketitle

\section{\label{sec:intro}Introduction}

Strong light-matter interactions provide powerful means to control and manipulate the physical and  chemical properties of matter. The latest advancements in laser-technology now enable the generation of few-cycle lasers in the IR and UV/Vis region with intensities of $\sim 10^{13}-10^{14}$ W cm$^{-2}$. At those intensities, the incident light can dramatically distort the electronic structure of bulk matter as the strength of the light-matter interaction becomes comparable to the strength of chemical bonds before the onset of dielectric breakdown. This opens exciting opportunities to create laser-dressed materials with structure-function relations that can be very different from those observed near thermodynamic equilibrium. Recent studies have demonstrated the creation of light-induced conical intersections \cite{Natan2016, Kuebel2020}, superconductivity \cite{Mitrano2016,Budden2021}, high harmonic generation \cite{Yoshikawa2017, UzanNarovlansky2022}, and light-wave electronics \cite{GarzonRamirez2020, Boolakee2022, Borsch2023}. To better understand  emerging  experiments, it is highly desirable to develop theoretical simulations based on realistic material Hamiltonians that go beyond parabolic bands or simple tight-binding models often employed to describe the properties of laser-dressed solids.

Theoretically, the external laser field can be considered as a time-periodic perturbation and, thus, can be treated exactly using Floquet theory. This has lead to a plethora of contributions in Floquet engineering \cite{Oka2019, Nuske2020, RodriguezVega2021, Earl2021,  Kobayashi2023, Barriga2024}, the study of the physical property of periodically driven systems. Importantly, recent observations have demonstrated that the Floquet picture remains accurate even for modeling the effects of ultrashort few-cycle lasers \cite{Lucchini2022, Ito2023}. 

A standard method of capturing the light-matter interactions in the Floquet engineering of solids is through the velocity gauge \cite{Faisal1997, Hsu2006, GomezLeon2013, Tiwari2023}, where interaction is bilinear in the vector potential of light ($\mathbf{A}(t)$) and the materials' momentum operator. This is in contrast to the length gauge where the interaction is bilinear in the electric field of light ($\mathbf{E}(t)$) and matter's dipole operator. Due to gauge invariance, both approaches provide identical results for the physical observables provided the basis is complete. The length gauge has the advantage of rapidly converging with the number of bands \cite{Virk2007, Yakovlev2017, Taghizadeh2017, Ventura2017}, but breaks the spatial periodicity of the solids. Its implementation leads to the Peierls substitution \cite{Schueler2021}  in tight-binding models  but often ignores intra-cell dipole transitions present in realistic solids. In turn, the velocity gauge has the advantage of respecting the space-periodicity of solids as needed to invoke Bloch theorem, and provides a key physical perspective of the laser-induced dynamics in solids \cite{Foeldi2017, Ernotte2018, Parker2019} and  computational advantages in  some cases \cite{Chirila2006, Dong2014, Yue2020, Yue22022, Kim2022, Mattiat2022, Xu2023}. However, it is limited by the fact that it usually requires a large number of bands for convergence \cite{Virk2007, Yakovlev2017, Taghizadeh2017}. Moreover, efficient and accurate first-principle electronic structure calculations for solids based on density functional theory (DFT) often require using the non-local pseudopotentials in the Hamiltonian. This pseudopotential approach leads to non-linear light-matter interaction terms in the velocity gauge \cite{Starace71, Girlanda81,  Beigi2001}, which are often ignored as they add significant computational burden especially when a  large number of bands are needed for convergence. Both these issues  make the computations in velocity gauge in the Floquet formalism using first-principle models approximate as, in practice, only a small finite number of bands can be tractably used to propagate the quantum dynamics.


In this paper, we propose an accurate and practical strategy to simulate the Floquet engineering of solids in the velocity gauge based on first-principle material Hamiltonians. For this, we first derive the light-matter interaction Hamiltonian in the velocity gauge from the length gauge taking into account the truncation of the Hilbert space. We refer to this scheme as the truncated velocity gauge. Our efforts are build upon Refs. \cite{Passos2018, Ventura2019}. We generalize these initial consideration to be able to capture realistic Hamiltonians and to drive-probe considerations where a strong pump laser drives the system out-of-equilibrium and a second weak laser probe its properties. The approach is motivated by the superior convergence properties of the length gauge in truncated Hilbert spaces. As shown, such space truncation within Floquet engineering leads to  terms that are non-linear in the vector potential of probe and drive laser expressed as a series of nested commutator between the position operator and the material Hamiltonian. The combined light-matter Hamiltonian in the truncated velocity gauge  respects the space-periodicity of the solid and allows the Bloch theorem to be invoked throughout. To capture the electronic structure, we employ generalized tight-binding models  constructed through Wannier interpolation \cite{Marzari2012} of first-principle calculations, a useful strategy  in describing linear and non-linear optical response of solids  \cite{Wang2017, Schueler2021, IbanezAzpiroz2022, Xu2023, Taghizadeh2018}. The Wannier function based approach allows us to compute the nested commutators up to all orders of the vector potential in  a straightforward way, even in presence of non-local pseudopotential terms in the Hamiltonian. This proposed strategy based on the truncated velocity gauge can be used to investigate the response properties of laser-dressed materials described through realistic Hamiltonians in a computationally efficient way that avoids the convergence issues inherent to the velocity gauge and the complications of computing the light-matter interactions with DFT-based solid Hamiltonians.

To exemplify and test the approach we use it to model the optical absorption properties of laser-dressed solids based on our previous work \cite{Tiwari2023} but now in the context of a first-principle realistic Hamiltonian for a solid. In this scenario, a crystal is dressed by a laser of arbitrary intensity and frequency, and the effective absorption properties of this laser-dressed system are then probed using a weak laser treated up to first-order in perturbation theory. We compute  the optical absorption spectra of laser-dressed \emph{trans}-polyacetylene (\emph{t}PA). The computations in the truncated velocity gauge with this realistic material Hamiltonian show faster convergence with respect to the number of bands than the usual velocity gauge, and accurately capture the spectrum even at high drive field strengths where the issues due to the non-local pseudopotential become important. Overall, our proposed strategy enables faster and accurate velocity gauge computations of light-matter interactions using a realistic description of the material Hamiltonians as needed in Floquet engineering.


This paper is organized as follows. In Sec. \ref{sec:theory} we derive the theory of laser-dressed solids in truncated Hilbert spaces and use it to characterize its optical response properties. In Sec. \ref{sec:Computation} we show how to  computationally implement the theory using a  realistic description of materials. In Sec. \ref{sec:results} we compare computations for \emph{t}PA in velocity and truncated velocity gauge with varying laser intensity and number of bands. We summarize our main  findings in Sec. \ref{sec:conclusion}.

\section{ Theory}\label{sec:theory}

\subsection{Velocity gauge Hamiltonian for truncated Hilbert spaces}

Inspired by Ref. \cite{Passos2018}, we now derive a Hamiltonian for a laser-driven solid in the velocity gauge starting from the length gauge while taking into account the effect due to the truncation of  Hilbert space. We opt to start with the  length gauge as it shows  much faster convergence with respect to the number of basis states compared to the velocity gauge \cite{Virk2007, Yakovlev2017, Taghizadeh2017}. 

In the length gauge, electrons in a solid satisfy the time-dependent Schr\"odinger equation (TDSE) \cite{Girlanda81, Cabellos2009, Taghizadeh2018}
\begin{align}
    \label{lengauge}
   i\hbar \frac{d }{d t}| \Psi_{l}(t) \rangle = ( \hat{H}_{0}  + e\mathbf{E}(t)\cdot \hat{\bm{r}}) | \Psi_{l}(t) \rangle  ,
\end{align}
where $| \Psi_{l}(t) \rangle$ is the many-particle electronic state in the length gauge, $\hat{H}_{0}$ is the space-periodic material Hamiltonian which can include contributions from both local and non-local potentials, $-e$ is the charge of electron, $\mathbf{E}(t)$ is the electric field due to the external laser and  $\hat{\bm{r}} = \sum_{i=1}^{M} \hat{\mathbf{r}}_{i} $  is the position operator of the $M$-electron system. We assume a uniform electric field in space and employ the electric dipole approximation, which is accurate as the unit cell is typically much smaller than the wavelength of light \cite{Chernyak1995}.

In the velocity gauge, the TDSE is  obtained by a Power-Zienau-Woolley transformation \cite{mukamel1995}
\begin{align}
|\Psi_{\mathrm{l}}(t)\rangle  = \hat{T}^{\dagger} (\hat{\bm{r}},t) |\Psi(t)\rangle  = e^{\frac{ie}{\hbar}\mathbf{A}(t)\cdot \hat{\bm{r}}}  |\Psi(t)\rangle 
\end{align}
of Eq. \eqref{lengauge}, where $|\Psi(t)\rangle$ is the state in velocity gauge and $\mathbf{A}(t)$ is the vector potential of light ($\mathbf{E}(t) =  - \frac{d \mathbf{A}(t)}{dt}$). We account for the truncation of Hilbert space by not assuming the canonical commutator relation  $[\hat{r}_{x},  \hat{P}_{y}] = i\hbar \hat{1} \delta_{xy}$,  where $\hat{\mathbf{P}} = \sum_{i=1}^{M} \hat{\mathbf{p}}_{i}$ is the true many-body momentum operator and $\hat{r}_{x},  \hat{P}_{y}$ are position and momentum along particular Cartesian coordinates of the many-body system. This implies that in the truncated Hilbert space $\hat{\mathbf{P}} \neq \frac{m_{e}}{i\hbar}[\hat{\mathbf{r}},\hat{H}_{0}]$ due to the Hilbert space truncation in addition to possible effects due to the non-local pseudopotential (see Sec. \ref{sec:Computation}).  This yields
\begin{equation}
i\hbar \frac{d }{d t}|\Psi (t)\rangle = \hat{H}(t)  |\Psi (t)\rangle ,
\end{equation}
where
\begin{align}
\nonumber
& \hat{H}(t)   =   \hat{H}_{0} + \left(\frac{e}{i\hbar}\right)  [ \mathbf{A}(t) \cdot \hat{\bm{r}}, \hat{H}_{0}  ] \\
\nonumber
& + \frac{1}{2!} \left(\frac{e}{i\hbar} \right)^2  [  \mathbf{A}(t) \cdot \hat{\bm{r}} ,  [  \mathbf{A}(t) \cdot \hat{\bm{r}} , \hat{H}_{0}  ]  ]  \\
\label{velgaugse}
&   + \frac{1}{3!} \left(\frac{e}{i\hbar} \right)^3  [  \mathbf{A}(t) \cdot \hat{\bm{r}}, [  \mathbf{A}(t) \cdot \hat{\bm{r}}, [  \mathbf{A}(t) \cdot \hat{\bm{r}}, \hat{H}_{0} ]  ]  ]  + \cdots  
\end{align}
is now the velocity gauge Hamiltonian in the truncated Hilbert space. Here, we have used the relation $e^{\hat{A}}\hat{B}e^{-\hat{A}}=\hat{B}+[\hat{A},\hat{B}]+\frac{1}{2!}[\hat{A},[\hat{A},\hat{B}]]+\frac{1}{3!}[\hat{A},[\hat{A},[\hat{A},\hat{B}]]]+\cdots$. In this work, we call this the ``truncated velocity gauge'' to differentiate it with the velocity gauge in complete Hilbert spaces. The nested commutator terms in Eq. \eqref{velgaugse} simplify to the regular velocity gauge when  the canonical commutator $[\hat{\mathrm{r}}_{x},  \hat{P}_{y}] = i\hbar \hat{1} \delta_{xy}$ is strictly satisfied and the material Hamiltonian does not include any non-local pseudopotential. In that case, $[ \hat{\bm{r}} , \hat{H}_{0} ] = \frac{ i \hbar}{ m_{e}} \hat{\mathbf{P}} $ leading to the second term in right side of Eq. \eqref{velgaugse} being  $e\hat{\mathbf{P}} \cdot \mathbf{A}(t)/m_{e}$, the third term being $(e\mathbf{A}(t))^2/2m_{e}$, while other higher order terms cancel.

A useful property of Eq. \eqref{velgaugse} is that it maintains the space-periodicity of the material Hamiltonian even in the presence of the external field. In fact, any spatial translation by integer multiples of the lattice vector $\mathbf{R}$ in Eq. \eqref{velgaugse} cancel out due to the commutator structure and the fact that $\hat{H}_{0}$ remains invariant under these translations. This enables to invoke Bloch theorem which will be used in Sec. \ref{sec:theorysecondquant}.

\subsection{Drive and probe laser considerations}
We extend the truncated velocity gauge Hamiltonian to simulate a physical situation in which an arbitrarily strong continuous wave (CW) laser dresses a solid while the effective non-equilibrium properties of this laser-dressed solid are probed using a weak CW laser. The net vector potential in such a situation is given by  $\mathbf{A}(t) = A_{\mathrm{d}} (t) \hat{\mathbf{e}}_{\mathrm{d}} + A_{\mathrm{p}} (t) \hat{\mathbf{e}}_{\mathrm{p}} $, where $ A_{\mathrm{d}} (t) $ and  $A_{\mathrm{p}}(t)$ represents the time-dependence of the drive and probe laser, respectively. In turn, $\hat{\mathbf{e}}_{\mathrm{d}}$ is the unit vector of the drive laser polarization while $\hat{\mathbf{e}}_{\mathrm{p}}$ is for the probe. Inserting this into Eq. \eqref{velgaugse} yields
\begin{align}
\nonumber
  & \hat{H}(t)  =  \hat{H}_{0} + \left(\frac{e}{i\hbar} \right)    [ \left( A_{\mathrm{d}} (t) \hat{\mathbf{e}}_{\mathrm{d}} + A_{\mathrm{p}} (t) \hat{\mathbf{e}}_{\mathrm{p}} \right) \cdot \hat{\bm{r}} , \hat{H}_{0}  ] \\
  \nonumber
  & + \frac{1}{2!} \left(\frac{e}{i\hbar} \right) ^2  [ \left( A_{\mathrm{d}} (t) \hat{\mathbf{e}}_{\mathrm{d}}  \right) \cdot \hat{\bm{r}}, [ \left( A_{\mathrm{d}} (t) \hat{\mathbf{e}}_{\mathrm{d}} + A_{\mathrm{p}} (t) \hat{\mathbf{e}}_{\mathrm{p}} \right) \cdot \hat{\bm{r}}, \hat{H}_{0} ]  ] \\
    \nonumber
  & + \frac{1}{2!} \left(\frac{e}{i\hbar} \right) ^2  [ \left( A_{\mathrm{p}} (t) \hat{\mathbf{e}}_{\mathrm{p}} \right) \cdot \hat{\bm{r}}, [ \left( A_{\mathrm{d}} (t) \hat{\mathbf{e}}_{\mathrm{d}} + A_{\mathrm{p}} (t) \hat{\mathbf{e}}_{\mathrm{p}} \right) \cdot \hat{\bm{r}}, \hat{H}_{0}  ]  ] \\
\label{separateterm}
  & + \cdots .
\end{align}

Equation \eqref{separateterm} contains terms proportional to  powers of the drive laser, probe laser and mixed vector potential terms. While we fully capture the effects of the drive, we only focus on the linear response to the probe. In this regime
\begin{align}
\label{parts}
  \hat{H}(t) \approx \hat{H}_{\mathrm{LD}}(t) + \hat{H}_{\mathrm{p}}(t) ,
  \end{align}
where $\hat{H}_{\mathrm{LD}}(t)$ is the laser-dressed Hamiltonian and $\hat{H}_{\mathrm{p}}(t)$  is the interaction due to the  probe laser to first order in $A_{\mathrm{p}}(t)$. The individual contributions can be simplified by writing them as
\begin{align}
  \nonumber
  \hat{H}_{\mathrm{LD}}(t) & =  \hat{H}_{0} + \left(\frac{eA_{\mathrm{d}}(t)}{i\hbar} \right)    [ \hat{\mathbf{e}}_{\mathrm{d}}  \cdot  \hat{\bm{r}}, \hat{H}_{0}  ]  \\
  \nonumber
 & + \frac{1}{2!} \left(\frac{eA_{\mathrm{d}} (t)}{i\hbar} \right)^2  [ \hat{\mathbf{e}}_{\mathrm{d}}  \cdot  \hat{\bm{r}} ,  [ \hat{\mathbf{e}}_{\mathrm{d}}  \cdot  \hat{\bm{r}} , \hat{H}_{0}  ]  ] + \cdots \\
\label{laserdressed}
& = \hat{H}_{0} + \hat{H}_{\mathrm{d}}(t)    ,
\end{align}
where the light-matter interaction with the driving laser is $\hat{H}_{\mathrm{d}}(t) = \sum_{j=1}^{\infty} \frac{1}{j!}\left( \frac{eA_{\mathrm{d}}(t)}{i\hbar} \right) ^{j} [ (\hat{\mathbf{e}}_{\mathrm{d}}  \cdot  \hat{\bm{r}} )^{j}, \hat{H}_{0} ] $ with  $ [ (\hat{\mathbf{e}}_{\mathrm{d}}  \cdot  \hat{\bm{r}} )^{j}, \hat{H}_{0}  ] \equiv \underbrace{ [  \hat{\mathbf{e}}_{\mathrm{d}}  \cdot  \hat{\bm{r}}, \hdots  [ \hat{\mathbf{e}}_{\mathrm{d}}  \cdot  \hat{\bm{r}}, \hat{H}_{0} ] \cdots   ]}_\text{$j$ times}$ being the $j^{\mathrm{th}}$-order nested commutators of position and material Hamiltonian. In turn, the interaction with the probe laser becomes:
\begin{align}
\nonumber
 \hat{H}_{\mathrm{p}}(t) & =   \frac{eA_{\mathrm{p}}(t)}{i\hbar}  
   \biggl( [  \hat{\mathbf{e}}_{\mathrm{p}}  \cdot \hat{\bm{r}} , \hat{H}_{0} ]  \\
 \nonumber
 &  +  \frac{1}{2!}\frac{eA_{\mathrm{d}}(t)}{i\hbar}  [ \hat{\mathbf{e}}_{\mathrm{p}}  \cdot \hat{\bm{r}} , [ 
 \hat{\mathbf{e}}_{\mathrm{d}}  \cdot \hat{\bm{r}}, \hat{H}_{0} ]]  \\ 
 \nonumber
 &  + \frac{1}{2!} \frac{e A_{\mathrm{d}}(t)}{i\hbar}  [ \hat{\mathbf{e}}_{\mathrm{d}}  \cdot \hat{\bm{r}} , [ 
 \hat{\mathbf{e}}_{\mathrm{p}}  \cdot \hat{\bm{r}}, \hat{H}_{0} ]]  + \cdots \biggr) \\
 \label{probeham}
 & =  \frac{e A_{\mathrm{p}}(t)}{m_{e}}    \hat{Z}(t) ,
\end{align}
where $\hat{Z}(t) $ is a truncated momentum operator that contains all the operator terms. Note that $\hat{Z}(t)$ contains powers of $A_{\mathrm{d}}(t)$ that arise due to the Hilbert space truncation. That is, space-truncation mixes the vector potential of the probe laser with that of the drive even at the level of the Hamiltonian.

The laser field due to the drive and the probe laser can be taken to be of any general time-periodic form and polarization. For simplicity, here we take the drive laser vector potential $ A_{\mathrm{d}}(t) = - \frac{E_{\mathrm{d}}}{\Omega}\sin(\Omega t) $, where  $E_{\mathrm{d}}$ is its amplitude and $\hbar \Omega$ its photon energy. The vector potential due to the probe laser is taken as $A_{\mathrm{p}} (t) = -\frac{E_{\mathrm{p}}}{\omega}\sin(\omega t) $ with amplitude  $E_{\mathrm{p}}$  and  photon energy $\hbar \omega$. Note  that  although the total Hamiltonian is not periodic in time due to the presence of the probe, the laser-dressed Hamiltonian $\hat{H}_{\mathrm{LD}}(t)$ is, with time period $T=\frac{2\pi}{\Omega}$.

\subsection{Optical response of a laser-dressed solid}
We quantify optical transitions in the laser-dressed non-equilibrium solid by the rate of change among laser-dressed states due to  interaction with the  probe laser  \cite{Mizumoto2006, Gu2018, Tiwari2023}. Mathematically, the rate  is given by
\begin{equation} 
\label{rateoftrans}
I(\omega)=\lim_{t\rightarrow\infty}\frac{W(t,\omega)}{t-t_{0}} ,
\end{equation}
where $W(t,\omega)$ is the probability of a probe photon of energy $\hbar \omega$ being absorbed or emitted in the laser-driven material after an interaction time $t-t_{0}$, with $t_{0}$ being the initial time. To first-order in the probe laser (confer Eq. (16) in Ref. \cite{Tiwari2023}) 
\begin{align}
\nonumber
I(\omega)  = & \lim_{t\rightarrow \infty}\frac{e^2 E_{\mathrm{p}}^{2}}{2\hbar ^{2} m_{e}^2\omega^2(t-t_0)}\iint_{t_{0}}^{t} dt_{1} dt_{2} C_{Z,Z}(t_1,t_2) \\
\label{rateofch}
& \times \mathrm{Re}[e^{-i\omega (t_1-t_2)}-e^{-i\omega( t_1+t_2)}] ,
\end{align}
where the operator $C_{Z,Z}(t_1,t_2)= \langle \Psi_a  | \hat{Z}_{\mathrm{I}}(t_1) \hat{Z}_{\mathrm{I}} (t_2)  | \Psi_a  \rangle $ represents the two-time correlation function of the truncated momentum operator $\hat{Z}(t)$ that couples to the probe laser and $|\Psi_{a}\rangle$ is the system's many-body state at $t_{0}$. In Eq. \eqref{rateofch}, we have adopted the interaction picture of  $\hat{H}_{\mathrm{LD}}(t)$ where $\hat{Z}_{\mathrm{I}}(t)=\hat{U}_{\mathrm{d}}^{\dagger} (t,t_0)  \hat{Z}(t)  \hat{U}_{\mathrm{d}}(t,t_0) $ and $\hat{U}_{\mathrm{d}}(t,t_0)=\mathcal{T}e^{\frac{-i}{\hbar}\int_{t_0}^{t}d\tau \hat{H}_{\mathrm{LD}}(\tau)}$ is the evolution operator of the laser-dressed system. Equation \eqref{rateofch} reduces to Eq. (16) in Ref. \cite{Tiwari2023} when the Hilbert space becomes complete and to the equilibrium theory of optical absorption \cite{Haug2009} in the absence of the drive laser $(E_{\mathrm{d}} = 0)$. While formally exact, Eq. \eqref{rateofch} is difficult to solve numerically as it requires propagating the system forward and backward in time for each pair of time $t_{1}$ and $t_{2}$, and for each value of $\omega$. To make progress, below we invoke Floquet theorem.

\subsection{Introducing second quantization}\label{sec:theorysecondquant}

We consider solids that can be described using an effective non-interacting Hamiltonian as constructed from DFT. In this case, the laser-dressed Hamiltonian Eq. \eqref{laserdressed} can be rewritten in second quantization as
\begin{align}
\label{manypartham}
\hat{H}_{\mathrm{LD}}(t) & = \sum_{\mathbf{k} \in \mathrm{BZ}}\sum_{u,v} \langle \psi_{u\mathbf{k}}|\mathcal{\hat{H}}_{\mathrm{LD}}(t)|\psi_{v\mathbf{k}}\rangle \hat{c}^{\dagger}_{u\mathbf{k}}\hat{c}_{v\mathbf{k}} ,
\end{align}
where  the creation (annihilation) operator $\hat{c}^{\dagger}_{u\mathbf{k}} (\hat{c}_{u\mathbf{k}})$ creates (annihilates) a fermion in  Bloch state $|\psi_{u \mathbf{k}} \rangle$ with band index $u$ and crystal momentum $\mathbf{k}$ in the first Brillouin zone (BZ).  The effective single-particle Hamiltonian
\begin{align}
\nonumber
\mathcal{\hat{H}}_{\mathrm{LD}}(t) & = \hat{h}_{0} + \sum_{j=1}^{\infty} \frac{1}{j!}\left( \frac{eA_{\mathrm{d}}(t)}{i\hbar}    \right) ^{j}  [ ( \hat{\mathbf{e}}_{\mathrm{d}}  \cdot  \hat{\mathbf{r}}) ^{j}, \hat{h}_{0} ] \\
\label{hamlighdress}
& = \hat{h}_{0} + \hat{h}_{\mathrm{d}}(t) ,
\end{align}
where $\hat{h}_{0}$ is the single-particle Hamiltonian of the pristine crystal, $\hat{\mathbf{r}}$ is the single electron position operator and $\hat{h}_{\mathrm{d}}(t) = \sum_{j=1}^{\infty} \frac{1}{j!} \left(\frac{eA_{\mathrm{d}}(t)}{i\hbar} \right) ^{j}  [ ( \hat{\mathbf{e}}_{\mathrm{d}}  \cdot \hat{\mathbf{r}}) ^{j}, \hat{h}_{0} ] $ is the laser-matter interaction with the driving field.

Since Eq. \eqref{hamlighdress} is periodic in space, we can invoke Bloch theorem. The $\mathbf{R}$-periodic Bloch modes  $| u \mathbf{k} \rangle  = \sqrt{V} e^{- i \mathbf{k} \cdot \hat{\mathbf{r}}}  | \psi_{u \mathbf{k}} \rangle $ can be obtained by solving the eigenvalue relation $ (e^{- i \mathbf{k} \cdot \hat{\mathbf{r}}} \hat{h}_{0} e^{ i \mathbf{k} \cdot \hat{\mathbf{r}}} ) | u \mathbf{k} \rangle = \epsilon_{u\mathbf{k}} | u \mathbf{k} \rangle $, where $\epsilon_{u \mathbf{k}}$ is the band energy and $V$ the volume of the crystal. Bloch theorem is useful in simplifying  the matrix elements in Eq. \eqref{manypartham} from the bulk to the single unit cell as
\begin{eqnarray}
\label{decoupled}
    \langle \psi_{u\mathbf{k}}| [ \hat{\mathbf{r}}, \hat{\mathcal{O}} ]   |\psi_{v\mathbf{k'}}\rangle = \delta_{\mathbf{k},\mathbf{k'}} N  \langle \psi_{u\mathbf{k}}| [ \hat{\mathbf{r}}, \hat{\mathcal{O}} ]  |\psi_{v\mathbf{k'}} \rangle_{\mathrm{UC}} ,
\end{eqnarray}
where  $\hat{\mathcal{O}}$ represents a space-periodic gauge covariant operator (such as $\hat{h}_{0}, [ \hat{\mathbf{r}} , \hat{h}_{0} ]$, etc.), $N$ is the number of unit cells (or, equivalently, the number of crystal momentum vectors) and $\langle \cdots \rangle_{\rm{UC}}$ is an  integral over the unit cell only. Equation \eqref{decoupled} shows that the dynamics of the laser-dressed system in the truncated velocity gauge is decoupled in the reciprocal space and all transitions due to the probe laser happen without change in momentum (i.e. vertically), as already reflected in Eq. \eqref{manypartham}.

To solve Eq. \eqref{rateofch} via the two-time correlation function $C_{Z,Z} (t_1,t_2)$, we need to determine  $\hat{Z}_{\mathrm{I}}(t)$. In second quantization,  $\hat{Z}_{\mathrm{I}}(t) = \sum_{\mathbf{k} \in \mathrm{BZ}} \sum_{u, v} \langle \psi_{u \mathbf{k}} | \hat{z}(t) | \psi_{v\mathbf{k}}\rangle \hat{c}^{\dagger}_{u\mathbf{k} , \mathrm{I}} (t) \hat{c}_{v\mathbf{k},\mathrm{I}}(t)$ with 
\begin{align}
  \nonumber
    \hat{z} (t) = & \frac{m_{e}}{i\hbar} \biggl( [ \hat{\mathbf{e}}_{\mathrm{p}}  \cdot \hat{\mathbf{r}} , \hat{h}_{0} ] + \frac{1}{2!} \frac{eA_{\mathrm{d}}(t)}{i\hbar}   [ \hat{\mathbf{e}}_{\mathrm{p}}  \cdot \hat{\mathbf{r}} , [ 
 \hat{\mathbf{e}}_{\mathrm{d}}  \cdot \hat{\mathbf{r}}, \hat{h}_{0} ]] \\ 
 &  + \frac{1}{2!}  \frac{eA_{\mathrm{d}}(t)}{i\hbar} [  \hat{\mathbf{e}}_{\mathrm{d}}  \cdot \hat{\mathbf{r}} , [ 
 \hat{\mathbf{e}}_{\mathrm{p}}  \cdot \hat{\mathbf{r}}, \hat{h}_{0} ]]  + \cdots \biggr) 
 \end{align}
 being the single-particle truncated momentum operator of the laser-dressed solid in terms of the nested commutators.  Taking into account that for non-interacting particles \cite{Tiwari2023}, the interaction picture fermionic operators are determined by 
 \begin{equation}
 \label{annhil}
   \hat{c}_{u\mathbf{k},\mathrm{I}}(t)=\sum_{v}  ( \hat{\mathcal{U}}(t,t_{0}) )_{u\mathbf{k} , v\mathbf{k}} \hat{c}_{v\mathbf{k}} ,
 \end{equation}
 where $\mathcal{\hat{U}} (t,t_{0}) = \mathcal{T}e^{\frac{-i}{\hbar}\int_{t_0}^{t} d\tau \mathcal{\hat{H}}_{\mathrm{LD}}(\tau)} $ is the time-ordered single-particle evolution operator of the laser-dressed system, then 
 \begin{align}
\nonumber
\hat{Z}_{\mathrm{I}}(t)&= \sum_{\mathbf{k} \in \mathrm{BZ}}\sum_{u,v,r,s} \langle \psi_{r\mathbf{k}}| \mathcal{\hat{U}}^{\dagger} (t,t_{0}) |\psi_{u\mathbf{k}} \rangle \langle \psi_{u\mathbf{k}}| \hat{z}(t) |\psi_{v\mathbf{k}} \rangle  \\
 \label{epoper}
& \times \langle\psi_{v\mathbf{k}} | \mathcal{U}(t,t_{0}) | \psi_{s\mathbf{k}}\rangle  \hat{c}^{\dagger}_{r\mathbf{k}} \hat{c}_{s\mathbf{k}} .
\end{align}
The problem of determining Eq. \eqref{rateofch} via  Eq. \eqref{epoper} can be solved once the matrix elements of $\mathcal{\hat{U}} (t,t_{0})$ are determined. To do so, we invoke Floquet theory.

\subsection{Application of the Floquet-Bloch theory}
Since the  single particle Hamiltonian for laser-dressed solid is periodic in both space $\mathcal{\hat{H}}_{\mathrm{LD}} (\hat{\mathbf{r}} , t ) = \mathcal{\hat{H}}_{\mathrm{LD}} ( \hat{\mathbf{r}} + \mathbf{R} , t ) $ and time  $\mathcal{\hat{H}}_{\mathrm{LD}}(\hat{\mathbf{r}},t)=\mathcal{\hat{H}}_{\mathrm{LD}}(\hat{\mathbf{r}},t+T)$, the laser-dressed system satisfies both  Floquet \cite{Floquet1883, Sambe1973} and Bloch \cite{Haug2009} theorem. Thus, the  Floquet-Bloch states \cite{Faisal1997}
\begin{eqnarray}
\label{fbstate}
|\Psi_{\alpha \mathbf{k}}(t)\rangle = \frac{1}{\sqrt{V}} e^{-iE_{\alpha  \mathbf{k}} t/ \hbar} e^{ i\mathbf{k} \cdot \hat{\mathbf{r}} } | \Phi_{\alpha \mathbf{k}} (t) \rangle 
\end{eqnarray} 
are solutions to the TDSE
\begin{eqnarray}
\label{oneelcsch}
i\hbar \frac{\partial}{\partial t} |\Psi_{\alpha \mathbf{k}}(t)\rangle=\mathcal{\hat{H}}_{\mathrm{LD}}(t)|\Psi_{\alpha \mathbf{k}}(t)\rangle .
\end{eqnarray}  
Here, the  Floquet-Bloch mode $|\Phi_{\alpha \mathbf{k}}(t) \rangle$ with index $\alpha$ and crystal momentum $\mathbf{k}$ is a function that is periodic in both time and space  $\big( \langle \mathbf{r}| \Phi_{\alpha \mathbf{k}} (t) \rangle = \Phi_{\alpha \mathbf{k}}(\mathbf{r},t) = \Phi_{\alpha \mathbf{k}}(\mathbf{r},t+T) = \Phi_{\alpha \mathbf{k}}(\mathbf{r+R},t)  \big)$ and $E_{\alpha \mathbf{k}}$ is the corresponding quasienergy.

The Floquet-Bloch modes and quasienergies can be determined by solving the following eigenvalue relation in Sambe space \cite{Sambe1973} (the tensor product space of the regular Hilbert space and the space spanned by the $T$-periodic Floquet Fourier basis $\{ e^{in\Omega t} \}$ where $n \in \mathbb{Z}$).
\begin{eqnarray}
\label{flqham}
\mathcal{\hat{H}}_{F} (\mathbf{k},\hat{\mathbf{r}},t) | \Phi_{\alpha \mathbf{k}} (t) \rangle = E_{\alpha \mathbf{k}} | \Phi_{\alpha \mathbf{k}} (t) \rangle ,
 \end{eqnarray}
where $ \mathcal{\hat{H}}_{F}(\mathbf{k},\hat{\mathbf{r}},t)  =  e^{-i\mathbf{k} \cdot \hat{\mathbf{r}}} \mathbf{\mathcal{\hat{H}}}_{\mathrm{LD}}(t) e^{i\mathbf{k} \cdot \hat{\mathbf{r}}} - i\hbar \frac{\partial}{\partial t}  $ is the Floquet-Bloch Hamiltonian. The Floquet-Bloch modes are uniquely defined in a Floquet-Brillouin zone (FBZ) with the fundamental FBZ being $\frac{-\hbar \Omega}{2}<E_{\alpha \mathbf{k}}\leq\frac{\hbar \Omega}{2}$ for $\mathbf{k}$ in the first BZ of the crystal. 
 
The Floquet-Bloch modes can be further expanded in terms of the time-periodic Floquet Fourier basis and the set of Bloch modes as
\begin{equation}
\label{ansatz2}
|\Phi_{\alpha \mathbf{k}}(t) \rangle = \sum_{n=-n_{F} }^{n_{F}} \sum_{u} F_{\alpha \mathbf{k}}^{(nu)} e^{in\Omega t} | u\mathbf{k} \rangle ,
 \end{equation}
 where $2n_{F}+1$ are the number of Floquet Fourier basis states needed for convergence. Substituting Eq. \eqref{ansatz2} into \eqref{flqham}, and taking the inner product in the Sambe space (that is, left multiplying by $\frac{1}{T} \int_{0}^{T} dt \langle v\mathbf{k}|e^{-im\Omega t} $ ) yields 
\begin{equation}
\label{flqeig}
\sum_{n,u}\Gamma_{mv,nu,\mathbf{k}}F_{\alpha \mathbf{k}}^{(nu)}=E_{\alpha \mathbf{k}}F_{\alpha \mathbf{k}}^{(mv)} ,
\end{equation}
where
\begin{align}
\nonumber
 \Gamma_{mv, nu, \mathbf{k}} & =  \frac{1}{T} \int_{0}^{T} dt  \langle v\mathbf{k}| e^{-i \mathbf{k} \cdot \hat{\mathbf{r}}} \mathcal{\hat{H}}_{\mathrm{LD}}(t) e^{i \mathbf{k} \cdot \hat{\mathbf{r}}} | u\mathbf{k} \rangle  e^{i(n-m)\Omega t} \\ 
 \label{flqhammatrix}
 & + n\hbar\Omega \delta_{nm}\delta_{uv}.
\end{align}
Substituting Eq. \eqref{hamlighdress} into right hand side of Eq. \eqref{flqhammatrix}  yields
\begin{align}
\nonumber
  & \Gamma_{mv,nu,\mathbf{k}}  = (\epsilon_{u \mathbf{k}} + n\hbar\Omega ) \delta_{nm}\delta_{uv} \\
  \nonumber
  & +   \frac{1}{T}\int_{0}^{T} dt \langle v\mathbf{k}| e^{-i \mathbf{k} \cdot \hat{\mathbf{r}}} \hat{h}_{\mathrm{d}}(t)  e^{i \mathbf{k} \cdot \hat{\mathbf{r}}}    | u\mathbf{k} \rangle  e^{i(n-m)\Omega t}   \\
\nonumber
 &  = (\epsilon_{u \mathbf{k}} + n\hbar\Omega ) \delta_{nm}\delta_{uv}  \\
 \nonumber
 & + \sum_{j=1}^{\infty} \sum_{l=0}^{j}  \frac{1}{j!} \left(\frac{eE_{\mathrm{d}}}{2\hbar  \Omega} \right)^{j}  \binom {j}{l} (-1)^{l} \\
  \label{flqblochmatrix}
 & \times \langle v\mathbf{k}| e^{-i \mathbf{k} \cdot \hat{\mathbf{r}}}   [ ( \hat{\mathbf{e}}_{\mathrm{d}}  \cdot \hat{\mathbf{r}}) ^{j}, \hat{h}_{0} ]     e^{i \mathbf{k} \cdot \hat{\mathbf{r}}}    | u\mathbf{k} \rangle   \delta_{j-2l+n,m}     ,
\end{align}
where we have used the binomial expansion, and where $\binom jl=\frac{j!}{l!(j-l)!}$. Note that the Floquet Hamiltonian matrix elements no longer form a block-tridiagonal matrix as in the case of the usual velocity gauge in complete Hilbert space because of the truncation.

The Floquet-Bloch states define the single-particle evolution operator \cite{Shirley1965}
\begin{eqnarray}
\label{evolut}
\mathcal{\hat{U}}(t,t_{0})=\sum_{\mathbf{k} \in \mathrm{BZ},\alpha}|\Psi_{\alpha \mathbf{k}}(t)\rangle \langle \Psi_{\alpha \mathbf{k}}(t_0)| 
\end{eqnarray}
needed to calculate the two-time correlation function $C_{Z,Z}(t_1,t_2)$ via Eq. \eqref{epoper}. Equation \eqref{evolut} can be verified by noting that it satisfies the TDSE in Eq. \eqref{oneelcsch}, and $\mathcal{\hat{U}}(t_{0},t_{0}) = \hat{1}$.

\subsection{Optical absorption coefficient for the laser-dressed solid}

Substituting Eqs. \eqref{evolut}, \eqref{ansatz2} and \eqref{fbstate} into \eqref{epoper} we get
\begin{align}
\nonumber
 \hat{Z}_{\mathrm{I}}(t) & =  \frac{1}{V^2}
\sum_{\mathbf{k} \in \mathrm{BZ}}\sum_{u,v}\sum_{\alpha ,\beta}e^{iE_{\alpha \beta \mathbf{k}}(t-t_0)/\hbar}\langle u\mathbf{k}|\Phi_{\alpha \mathbf{k}}(t_{0})\rangle \\ 
\label{epoperlast}
& \times \langle\Phi_{\beta \mathbf{k}}(t_{0})| v\mathbf{k}\rangle \mathcal{Z}_{\alpha\beta \mathbf{k}}(t)  \hat{c}^{\dagger}_{u\mathbf{k}} \hat{c}_{v\mathbf{k}} ,
\end{align}
where we have taken into account the orthonormality of Bloch states $\langle \psi_{u\mathbf{k}} | \psi_{v\mathbf{k'}} \rangle =  \delta_{u v}\delta_{\mathbf{k} \mathbf{k'}}$ and, where $E_{\alpha \beta \mathbf{k}}=E_{\alpha \mathbf{k}}-E_{\beta  \mathbf{k}}$. Here, we define the truncated momentum matrix elements between the Floquet-Bloch modes $\alpha, \beta$ with crystal momentum $\mathbf{k}$ as
\begin{align}
\label{transitionmme}
\mathcal{Z}_{\alpha\beta \mathbf{k}}(t) & =  \frac{1}{V} \langle\Phi_{\alpha \mathbf{k}}(t)| e^{-i \mathbf{k} \cdot \hat{\mathbf{r}}} \hat{z}(t) e^{i \mathbf{k} \cdot \hat{\mathbf{r}}} | \Phi_{\beta \mathbf{k}}(t)\rangle .
\end{align}
Since the Floquet-Bloch modes and $\hat{z}(t)$  are $T$-periodic, $\mathcal{Z}_{\alpha\beta \mathbf{k}}(t)$  can be further expanded  in a  Fourier series
\begin{eqnarray}
\label{momperiodic}
\mathcal{Z}_{\alpha \beta \mathbf{k}}(t)=\sum_{n=-\infty}^{\infty}\mathcal{Z}_{\alpha\beta \mathbf{k}}^{(n)}e^{in\Omega t} ,
\end{eqnarray}
where 
\begin{equation}
    \label{fbmme}
\mathcal{Z}_{\alpha\beta \mathbf{k}}^{(n)}=\frac{1}{T}\int_{0}^{T}dt \mathcal{Z}_{\alpha\beta \mathbf{k}}(t)e^{-in\Omega t} 
\end{equation}
is the $n$-th Fourier component of the truncated momentum matrix element.

Following the procedure detailed in Ref. \cite{Tiwari2023}, we construct the two-time correlation function using Eqs. \eqref{epoperlast} and \eqref{momperiodic} and inserting it into Eq. \eqref{rateofch}. From this, we separate $I(\omega)$ into distinct contributions due to optical absorption and stimulated emission. The net rate of absorption $R(\omega)$ is obtained by subtracting the rate of stimulated emission from the rate of absorption. The optical absorption coefficient $A(\omega) = \frac{R(\omega)\hbar\omega}{VI_{0}}$ is obtained  as the ratio of the power absorbed by the incident probe laser per unit volume and incident light flux $I_{0} = \epsilon_{0} E_{\mathrm{p}}^2 c n_{r} / 2 $, where $\epsilon_{0} $ is the permittivity of vacuum,  $ c$ the speed of light, and $n_{r}$ the refractive index of the material \cite{Dresselhaus2018}. In this way, we obtain an  expression for the optical absorption coefficient of the laser-dressed solid using the truncated velocity gauge 
\begin{align}
\nonumber
A(\omega)    & =  \frac{e^{2}\pi}{ m_{e}^2 \epsilon_{0}c n_{r} V \omega}\sum_{\mathbf{k} \in \mathrm{BZ} }\sum_{\alpha, \beta}\sum_{n=-\infty}^{\infty}\Lambda_{\alpha \beta \mathbf{k}}|\mathcal{Z}_{\alpha\beta \mathbf{k}}^{(n)}|^2  \\
\label{final}
&  \times   [\delta(E_{\alpha\beta \mathbf{k}}+n\hbar\Omega-\hbar \omega)-\delta(E_{\alpha\beta \mathbf{k}}+n\hbar\Omega+\hbar \omega)] ,
\end{align}
where each of the terms in the sum are the contributions due to a particular optical transition from Floquet-Bloch mode $\beta \rightarrow \alpha$ at crystal momentum $\mathbf{k}$.  Here, 
\begin{align}
\nonumber
\Lambda_{ \alpha \beta \mathbf{k} }& = \frac{1}{V^4} \sum_{u', u}|\langle u\mathbf{k}|\Phi_{\beta \mathbf{k}}(t_{0})\rangle |^2 |\langle\Phi_{\alpha \mathbf{k}}(t_{0})|u'\mathbf{k}\rangle |^2 \\
\label{popfact}
& \times \bar{n}_{u \mathbf{k}}(1-\bar{n}_{u' \mathbf{k}}) 
\end{align}
is the so-called population factor that ensures that the initial state with label $\beta \mathbf{k}$ is occupied and final state $\alpha \mathbf{k}$ is empty, and  $\bar{n}_{u \mathbf{k}} = \langle \Psi_{a} | \hat{c}^{\dagger}_{u\mathbf{k}} \hat{c}_{u\mathbf{k}} | \Psi_{a} \rangle $ represents the thermal occupation number of band $u$ and crystal momentum $\mathbf{k}$ at preparation time.

Equation \eqref{final} shows that the optical absorption in a laser-dressed solid is akin to that of the equilibrium response theory \cite{Haug2009}  but with the  Floquet-Bloch modes playing the role of pristine eigenstates as the optical absorption is seen to emerge from transitions among them. The Bohr transition energy is given by the difference in the quasienergy of the participating modes ($E_{\alpha \beta \mathbf{k}}$) along with the $n\hbar\Omega$ term which corresponds to the number of Floquet-Brillouin zones separating them. The first term in Eq. \eqref{final} represents absorption; the second, stimulated emission. The intensity of a transition from $\beta \rightarrow \alpha$ separated by $n$ FBZ is determined by the population factor $\Lambda_{\alpha \beta \mathbf{k}}$ and the $n$-th Fourier component of the truncated momentum matrix element $\mathcal{Z}_{\alpha\beta \mathbf{k}}^{(n)}$. The population factor guarantees that an optical transition happens from an initially occupied band to an empty one. Overall, this shows that the Floquet-Bloch modes are the natural states to understand the optical absorption properties of laser-dressed solids.

The theory presented here is valid for realistic material Hamiltonians as obtained from first-principle based computations. It differs from our previous theory in the velocity gauge \cite{Tiwari2023} in that it takes into account Hilbert space truncations inherent to any electronic structure computation. Specifically, this changes the laser-dressed Hamiltonian in Eq. \eqref{flqhammatrix} used to obtain the Floquet-Bloch modes using Eq. \eqref{flqeig} and the transition matrix elements of the new truncated momentum operator coupled to the probe laser, Eq. \eqref{transitionmme}, which now need to include the driving laser. The Floquet-Bloch Hamiltonian Eq. \eqref{flqblochmatrix} in the truncated Hilbert space is not block tridiagonal anymore as the nested commutator terms depend on the powers of the drive vector potential (see right side of Eq. \eqref{flqblochmatrix}). This means that the Floquet-Bloch Hamiltonian constructed in the truncated velocity gauge is less sparse and would comparatively require more Floquet Fourier basis states ($n_{F}$) for convergence compared to the Floquet-Bloch Hamiltonian in the usual velocity gauge. This theory reduces to that in Ref. \cite{Tiwari2023} when the basis is complete  such that  $[\hat{\mathrm{r}}_{i},  \hat{P}_{j}] = i\hbar \hat{1} \delta_{ij}$ and the material Hamiltonian does not include any non-local potential.

\section{Computational method \label{sec:Computation}}

The computation of the optical absorption spectrum of laser-dressed solids for the complete Hilbert space \cite{Tiwari2023} and truncated spaces (Sec. \ref{sec:theory}) requires the single-particle laser-dressed Hamiltonian in the Bloch state basis to construct the Floquet-Bloch Hamiltonian and obtain the Floquet-Bloch modes. We now discuss how this is accomplished in both complete and truncated Hilbert spaces using  a realistic  Hamiltonian for the material and its integration into \textsc{FloqticS} (Floquet optics in Solids) which is a computational package to characterize the optical properties of laser-dressed solids \cite{code23}.

\subsection{Complete Hilbert space \label{sec:computationvelocity}}
As a benchmark for the truncated velocity gauge (Sec. \ref{sec:theory}), we perform computations assuming a complete Hilbert space as detailed in Ref. \cite{Tiwari2023}. The solid Hamiltonian is based on DFT as obtained from first-principle computational packages (such as \textsc{Quantum ESPRESSO} \cite{Giannozzi2017}). When employing a realistic DFT Hamiltonian in Floquet engineering, two challenges emerge: (i) basis-set convergence, and (ii) the modification of the light-matter interactions due to non-local pseudopotential ($V_{\mathrm{NL}}$). The non-local pseudopotentials are used to eliminate  contribution due to inert core electrons in the electronic structure \cite{Pickett1989, Vanderbilt1990} and enables efficient electronic structure calculations  compared to all electron methods. However, as generalized by Louie et al. \cite{Beigi2001} and described in previous work \cite{Girlanda81,Starace71}, the single-particle Hamiltonian of a solid in the presence of $V_{\mathrm{NL}}$ interacting with a strong driving laser in  dipole approximation  is 
\begin{align}
\nonumber
    \mathcal{\hat{H}}_{\mathrm{LD}}^{\mathrm{DFT}}(t) & =  \hat{h}_{0} - \frac{e\mathbf{A}_{\mathrm{d}}(t)}{m_{e}} \cdot \bigl( \hat{\mathbf{p}} - \frac{i m_{e}}{\hbar}[\hat{\mathbf{r}} , \hat{V}_{\mathrm{NL}} ] \bigr)  \\ 
    \label{louienonlocalham}
    & + \frac{e^2 \mathbf{A}_{\mathrm{d}}^2(t)}{2m_{e}} + \frac{1}{2!} \biggl( \frac{ie\mathbf{A}_{\mathrm{d}}(t) }{\hbar}\biggr)^2 [\hat{\mathbf{r}} ,[\hat{\mathbf{r}} , \hat{V}_{\mathrm{NL}}]] + \cdots .
\end{align}
Equation  \eqref{louienonlocalham} simplifies to the usual velocity gauge coupling $\hat{\mathbf{p}}\cdot \mathbf{A}_{\mathrm{d}}(t)$ in the absence of the non-local pseudopotential. In our computations, we consider the single-particle laser-dressed Hamiltonian
\begin{align}
\label{nonlocalham}
& \mathcal{\hat{H}}_{\mathrm{LD}}^{\mathrm{(VG)}}(t) = \hat{h}_{0} - \frac{e\mathbf{A}_{\mathrm{d}}(t)}{m_{e}} \cdot \bigl( \hat{\mathbf{p}} - \frac{i m_{e}}{\hbar}[\hat{\mathbf{r}} , \hat{V}_{\mathrm{NL}} ] \bigr) 
\end{align}
obtained from Eq. \eqref{louienonlocalham} by only considering the linear coupling to the drive vector potential, and applying a gauge transformation to remove the purely time-dependent term proportional to $\mathbf{A}_{\mathrm{d}}^2(t)$ (this term only contributes to an overall phase to the wavefunction with no observable consequences). Computing the matrix elements of Eq. \eqref{nonlocalham} is directly accessible in \textsc{Quantum ESPRESSO} \cite{Kageshima1997, Tobik2004} while the neglected non-linear coupling terms due to  $V_{\mathrm{NL}}$ are not. This approximation implies that the calculations involving Eq. \eqref{nonlocalham} can become inaccurate for strong driving laser. As described in Sec. \ref{sec:computationtruncated}, by using the maximally-localized Wannier functions (MLWFs) \cite{Marzari2012}, it is technically possible to capture the non-linear interaction terms in Eq. \eqref{louienonlocalham}. However, such an approach remains, in practice, impractical as it involves determining the MLWFs and the non-linear coupling matrix elements for the large number of bands required for convergence in the velocity gauge formalism.

The  methodology to calculate the optical absorption spectra of laser-dressed solids in the velocity gauge is detailed in our previous work \cite{Tiwari2023} and is  implemented in \textsc{FloqticS} \cite{code23}. The code requires as input the band structure, light-matter interaction  matrix elements (second term in Eq. \eqref{nonlocalham}) in Bloch states basis, $\mathbf{k}$-vectors sampling the BZ and the drive laser parameters as input. It outputs the absorption spectrum, intensity of absorption as a function of $\hbar\omega$, obtained by broadening the absorption lines using a Lorentzian function of a chosen width.

\subsection{Truncated Hilbert space \label{sec:computationtruncated} }

Determining the Floquet state through Eq. \eqref{flqeig} and calculating the optical absorption spectrum of laser-dressed solid using Eq. \eqref{final} in the truncated velocity gauge requires evaluating the  nested commutators of  position and material Hamiltonian in the Bloch state basis. To do so,  it is convenient to describe the  Hamiltonian using a general tight-binding description constructed from the MLWFs as obtained through Wannier90 \cite{Pizzi2020}. In this case, a Bloch-like state in the Wannier gauge can be expressed as \cite{Pedersen2001, Marzari2012, Wang2007, Silva2019}
\begin{equation}
| \psi_{f \mathbf{k}}^{(\mathrm{W})}\rangle = \frac{1}{\sqrt{N}} \sum_{\mathbf{R}} e^{i \mathbf{k} \cdot \mathbf{R}} | f \mathbf{R} \rangle ,
\end{equation}
where $| f \mathbf{R} \rangle $ is the $f$ Wannier function localized in the unit cell at position $\mathbf{R}$. We assume that these Wannier functions form a complete ($\sum_{f \mathbf{R}} | f \mathbf{R} \rangle \langle f \mathbf{R} | = \hat{1}$) and orthonormal ($\langle f \mathbf{R'} | f' \mathbf{R}' \rangle = \delta_{ff'}\delta_{\mathbf{RR'}} $) basis. The completeness relation requires a convergence check on the number of MLWFs used. The material Hamiltonian  expressed in these Bloch-like states is given by 
\begin{equation}
    [h^{(\mathrm{W})}_{\mathbf{k}}]_{fg} = \langle  \psi_{f \mathbf{k}} ^{(\mathrm{W})} |  \hat{h}_{0} | \psi_{g \mathbf{k}} ^{(\mathrm{W})} \rangle = \sum_{\mathbf{R}} e^{i \mathbf{k} \cdot \mathbf{R}} t_{f\mathbf{0}g\mathbf{R}} ,
\end{equation}
where $t_{f\mathbf{0}g\mathbf{R}} = \langle f \mathbf{0} | \hat{h}_{0} | g \mathbf{R} \rangle $ is the tight-binding hopping parameters. The Wannier interpolated band structure can be obtained by diagonalizing $h^{(\mathrm{W})}_{\mathbf{k}}$ that is,
\begin{equation}
\label{tightbindingband}
  [U_{ \mathbf{k}  } ^{\dagger}   h^{(\mathrm{W})}_{\mathbf{k}}  U_{\mathbf{k} } ]_{uv} =  \epsilon_{u \mathbf{k}} \delta_{u v}  ,
\end{equation}
where $U_{\mathbf{k} }$ is the unitary operator that leads to the diagonalization.

The true Bloch eigenstates of the material Hamiltonian can be expanded in the $ \{| \psi_{f \mathbf{k}} ^{(\mathrm{W})} \rangle \}$ basis as
\begin{equation}
\label{blochstatewannier}
|\psi_{u\mathbf{k}} \rangle = \sum_{f }  | \psi_{f \mathbf{k} } ^{(\mathrm{W})}\rangle U_{uf \mathbf{k}  } .
\end{equation} 
We use this expansion to evaluate the space-periodic nested commutator as 
\begin{align}
\nonumber
  & \frac{1}{i} \langle \psi_{v\mathbf{k}}|      [  \hat{\mathbf{r}} , \hat{\mathcal{O}}  ]   | \psi_{u\mathbf{k}} \rangle    = \frac{1}{i} \sum_{f ,f' } U_{vf \mathbf{k}  }^{\dagger} \langle \psi_{f \mathbf{k}} ^{(\mathrm{W})} |  [ \hat{\mathbf{r}} , \hat{\mathcal{O}}  ] | \psi_{f' \mathbf{k}} ^{(\mathrm{W})}\rangle U_{uf' \mathbf{k}  }  \\
\label{commutmme}
 & = \sum_{f ,f' } U_{vf \mathbf{k} }^{\dagger} \bigl( \nabla_{\mathbf{k}} \mathcal{O}_{f f'}^{(\mathrm{W})}(\mathbf{k}) - i  [ \hat{A} , \hat{\mathcal{O}}  ]_{f f' \mathbf{k}} \bigr) U_{uf' \mathbf{k} } ,
\end{align}
where $\hat{\mathcal{O}}_{f f'}^{(\mathrm{W})}(\mathbf{k}) \equiv \langle f \mathbf{k} ^{(\mathrm{W})} |  \hat{\mathcal{O}}  | f' \mathbf{k} ^{(\mathrm{W})}\rangle = \sum_{\mathbf{R}} e^{i\mathbf{k} \cdot \mathbf{R}} \langle f \mathbf{0} | \hat{\mathcal{O}}| f' \mathbf{R} \rangle$, and  $A_{ ff'}( \mathbf{k}) \equiv \sum_{\mathbf{R}} e^{i\mathbf{k} \cdot \mathbf{R}} d_{f \mathbf{0} f' \mathbf{R}} $ is the Fourier-transformed position operator with $d_{f \mathbf{R}' g \mathbf{R}} =  \langle f \mathbf{R}' | \hat{\mathbf{r}} | g \mathbf{R} \rangle$.  We provide a detailed derivation of Eq. \eqref{commutmme} in terms of the Wannier functions in the Appendix.

Equation \eqref{commutmme} is useful to exactly capture the light-matter interaction as all the nested commutator terms in Eq. \eqref{flqhammatrix} are of this form. As shown, $\langle \psi_{v\mathbf{k}}|      [  \hat{\mathbf{r}} , \hat{\mathcal{O}}  ]   | \psi_{u\mathbf{k}} \rangle $ require the matrix elements of  $\hat{\mathcal{O}}$ in the Bloch-like state basis. Thus, every next-order commutator term (for example $[\hat{\mathbf{r}}, [\hat{\mathbf{r}} , \hat{\mathcal{O}}]]$) requires the matrix elements of the previous-order commutator ($[\hat{\mathbf{r}}, \hat{\mathcal{O}}]$). This property of Eq. \eqref{commutmme} allows us to obtain any general order matrix elements iteratively starting with $\hat{\mathcal{O}} = \hat{h}_{0}$. Furthermore, evaluating Eq. \eqref{commutmme} using the MLWF allows us to go beyond previous works \cite{Passos2018} where only the first few terms were required in the nested commutators. The computations performed using Eq. \eqref{commutmme} can accommodate both the intercell and the intracell position operator matrix elements among the Wannier functions present in realistic materials. Hence, it goes beyond the length gauge based Floquet engineering implemented using Peierls substitution that often ignores the intra-cell dipole matrix elements. Further note that the procedure remains valid even in presence of the non-local pseudopotentials as the derivation of Eq. \eqref{commutmme} does not impose any restriction on the form of $\hat{h}_{0}$ other than it being space-periodic and described using the generalized tight-binding models.

Note that the computations in the truncated velocity gauge enables exact treatment of the drive and probe laser up to all orders [using Eq. \eqref{commutmme}]. This is in contrast to the approximate treatment in velocity gauge for DFT-based material Hamiltonian  [Eq. \eqref{nonlocalham}] assuming a complete Hilbert space.

\subsection{Integration into \textsc{FloqticS}}
The truncated velocity gauge computations for Floquet engineering has been computationally implemented into  \textsc{FloqticS} \cite{code23} which is freely available through \textsc{GitHub}. The code allows the efficient calculation of the optical absorption spectrum of a solid that is driven by a laser of arbitrary intensity and frequency, interfaces with electronic structure codes for space-periodic materials, and is fully parallelized.

The code takes the $\mathbf{k}$-vector sampling the BZ, tight-binding parameters ($t_{f\mathbf{0}g\mathbf{R}} $ and $d_{f\mathbf{0}g\mathbf{R}}$) describing the realistic material and the drive laser parameters as inputs. The $\mathbf{k}$-vectors can be obtained by imposing the Born-Von-Karman boundary condition using the lattice vectors of the material while the parameters $t_{f\mathbf{0}g\mathbf{R}} $ and $d_{f\mathbf{0}g\mathbf{R}}$, for a given number of Wannier functions taken in a unit cell, are obtained from  \textsc{Wannier90} \cite{Pizzi2020}. The code first computes the $\epsilon_{u \mathbf{k}}$ using Eq. \eqref{tightbindingband} and the specified number of nested commutator matrix elements using Eq. \eqref{commutmme} for the given probe and drive laser polarization $\hat{\mathbf{e}}_{\textrm{d}}, \hat{\mathbf{e}}_{\textrm{p}}$ for each $\mathbf{k}$-vector in the BZ. The code then constructs and diagonalizes the Floquet-Bloch Hamiltonian using Eqs. \eqref{flqeig} and \eqref{flqhammatrix} to obtain $E_{\alpha \mathbf{k}}$ and the coefficients $F_{\alpha \mathbf{k}}^{(nu)}$ for the given drive laser amplitude and photon energy. Computations should be checked for convergence on the number of nested commutator terms in Eq. \eqref{flqeig}, the Floquet Fourier basis states ($n_{F}$), and the number of Wannier functions. The code proceeds to calculate the Fourier components of the truncated momentum matrix elements Eq. \eqref{transitionmme}, and the population factor Eq. \eqref{popfact} using $\bar{n}_{u\mathbf{k}}$ as inputs. In the end, the code outputs the optical absorption spectrum of the laser-dressed material with each absorption line broadened using a Lorentzian function of a given width.

\textsc{FloqticS} uses the highly parallelized direct diagonalization package \textsc{ELPA} (Eigenvalue soLvers for Petaflop Applications) \cite{Marek2014} for the diagonalization of the Floquet-Bloch Hamiltonian. It also parallelizes the calculation of the Fourier components of the truncated momentum matrix elements by distributing the components across different processors. The efficient parallelization of the computation allows us to compute the absorption properties with a finer Brillouin zone sampling of a realistic solid in a tractable computational time.

\section{Results \label{sec:results}}

To illustrate and test the methodology, we focus on the optical absorption spectrum of the laser-dressed \emph{trans}-polyacetylene (\emph{t}PA) as it provides a realistic model system that is simple enough due to its one-dimensionality to enable us to  check the convergence of the absorption spectrum with respect to modeling parameters in tractable computational time. We contrast the absorption coefficient as computed in velocity (Sec. \ref{sec:computationvelocity}) and truncated velocity  (Sec. \ref{sec:computationtruncated}) gauge and their convergence properties. Because of the approximations in Eq. \eqref{nonlocalham},  the velocity gauge is only expected to be accurate for weak to moderate strengths of the laser-dressing. By contrast, the truncated velocity gauge is expected to be accurate for arbitrary laser strength.

The qualitative features of the optical absorption of laser-dressed matter have already been discussed in Ref. \cite{Tiwari2023}. Briefly, the findings suggest that when a periodic solid is driven out of equilibrium by a continuous wave laser, the Floquet-Bloch states are formed. These laser-dressed states are replicas of the valence and conduction band that are separated from each other by integer multiples of the drive photon energy. When this laser-dressed band structure is probed, it leads to emergence of the below-band edge transitions, absorption sidebands and mid-infrared frequency absorption/stimulated emission features. The below-band edge features and absorption sidebands occur due to transitions among the Floquet-Bloch replicas while the low-frequency transitions emerge due to the hybridization of the Floquet-Bloch states. Overall, a strong non-resonant laser can substantially transform the optical absorption spectrum of a solid.

In the following computations, for definitiveness, we focus on dressing with non-resonant drive laser with photon energy $\hbar\Omega=0.4$ eV.  Throughout, the probe and drive laser is taken to be linearly polarized with polarization direction along the lattice growth direction. For all our calculations below, we obtain converged results with $n_{F}=150$, use a Lorentzian lineshape for the absorption lines with 0.04 eV width and remove transition below 0.06 eV to account for the discrete sampling of the Brillouin zone.

\subsection{Electronic structure and its Wannier interpolation}

\begin{figure}[htbp!]
    \centering
    \includegraphics[width=0.48\textwidth]{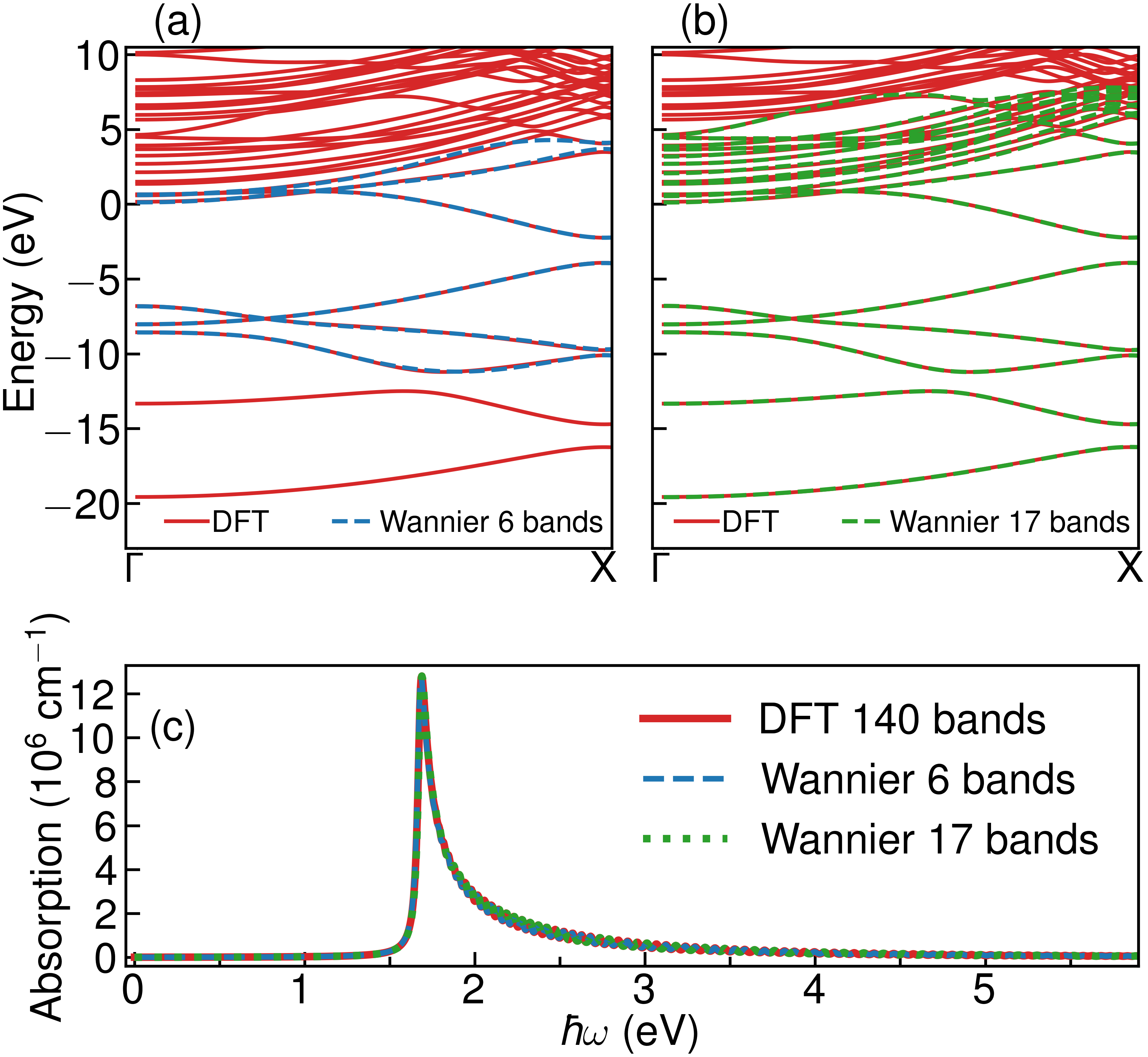}
    \caption{(a)-(b) Band structure of the \emph{t}PA obtained from DFT (red lines) along the $\Gamma \rightarrow $ X direction in the BZ. Dashed lines are interpolation with (a) 6- (blue) and (b) 17- (green) Wannier functions   per unit cell. (c) Equilibrium absorption spectrum of \emph{t}PA computed using DFT in the velocity gauge (red), and truncated velocity gauge from the 6-band (blue) and 17-band (green) generalized tight-binding models. }
    \label{fig:plot1}
\end{figure}

The first-principle self-consistent field computations of the electronic structure of \emph{t}PA are done in \textsc{Quantum ESPRESSO}, using the local density approximation (LDA) with the Perdew-Zunger parametrization for the exchange-correlation functional \cite{Perdew1981}, ultra-soft pseudopotential  \cite{DalCorso2014}, and a plane wave cutoff of 100 Ry that yield   160 converged  bands. We use the geometry of \emph{t}PA from Ref. \cite{Ferretti2012} with bond length alteration of 1.34/1.54 \r{A} and unit cell of dimensions $2.496 \times 10 \times 10$ \r{A}$^{3}$. The obtained DFT based band structure is shown in Fig. \ref{fig:plot1}(a)-(b) (red lines). The Brillouin zone is discretized using $500 \times 1\times 1$ $\mathbf{k}$-vector grid. The obtained band gap of 1.67 eV at X  point is in the range of experimentally observed band gaps \cite{Leising88,Fincher79}.  Figure \ref{fig:plot1} also shows the band structure constructed from the 6 (Fig. \ref{fig:plot1}(a), blue lines) and 17 (Fig. \ref{fig:plot1}(b), green lines) Wannier functions tight-binding models and their comparison with the DFT band structure. In both cases, the Wannier functions accurately interpolate the DFT results. We also contrast the equilibrium optical absorption spectrum of the \emph{t}PA in Fig. \ref{fig:plot1}(c) computed with the tight-binding model to that calculated using 140 bands in DFT. The equilibrium optical absorption spectrum shows an absorption edge at the direct band gap of 1.67 eV and subsequent decrease in $A(\omega)$ as expected for one-dimensional solids \cite{Haug2009}. As seen, the equilibrium absorption spectrum matches for the three different  Hamiltonians. Overall, this shows that the Wannier interpolation is highly accurate and can be further used to simulate the laser-dressed properties.

\subsection{Optical absorption spectrum of laser-dressed materials in velocity gauge}\label{sec:resultsvelocity}

\begin{figure}[htbp!]
    \centering
    \includegraphics[width=0.48\textwidth]{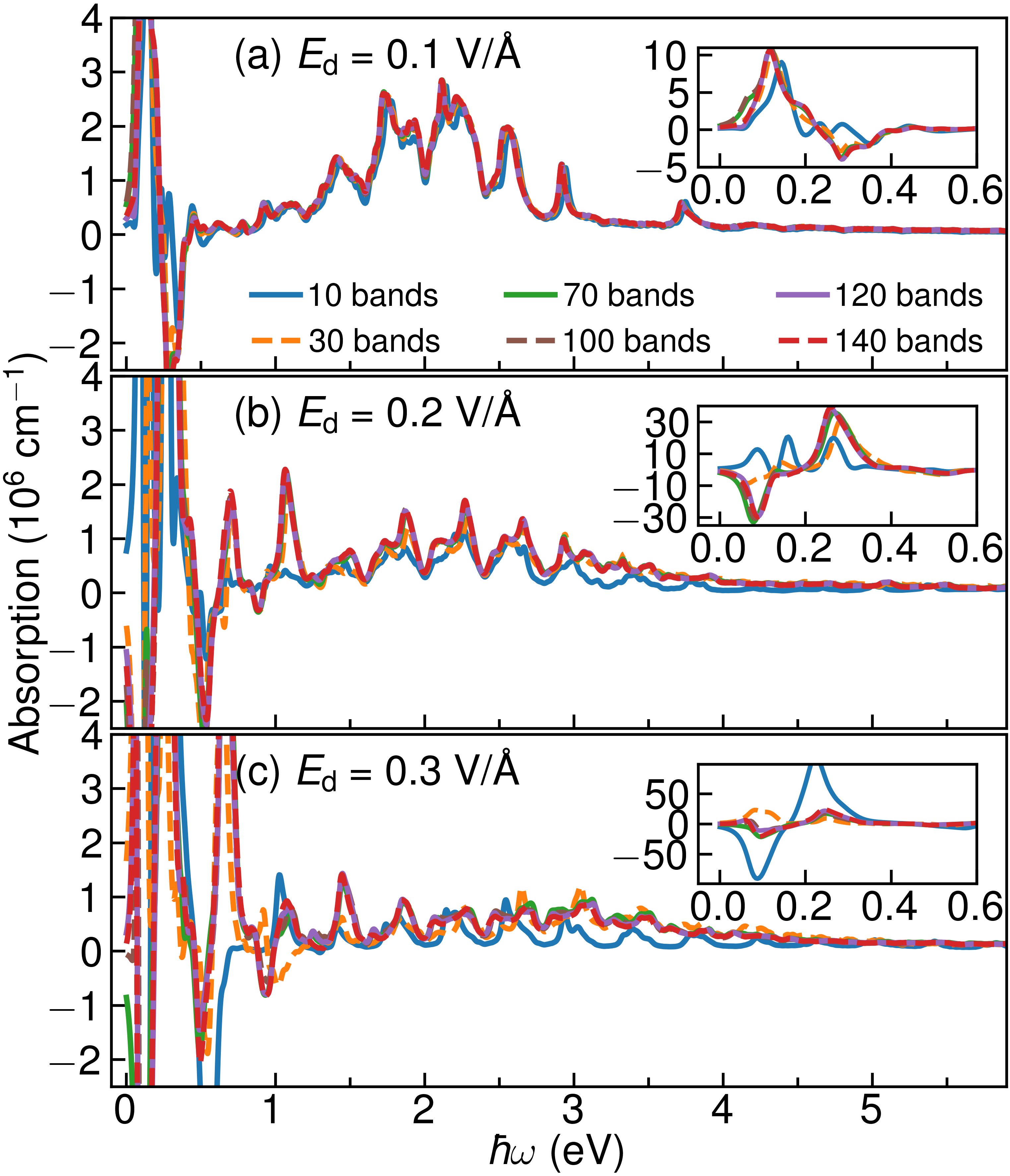}
    \caption{Optical absorption spectrum of \emph{t}PA calculated using Eq. \eqref{nonlocalham} in the velocity gauge for $E_{\mathrm{d}}$  (a) 0.1 V/\r{A} (b) 0.2 V/\r{A} and (c) 0.3 V/\r{A}. Different lines correspond to the different number of bands taken in the computation. Insets show the spectrum in the low-frequency region.}
    \label{fig:plot2}
\end{figure}

We obtain the absorption spectrum for the $500\times1\times1$ $\mathbf{k}$-vector grid in the Brillouin zone for varying drive laser amplitude $E_{\mathrm{d}} = 0.1-0.3$ V/\r{A}.  The Fermi energy of the system is at -3.065 eV such that bands below (above) this energy are valence (conduction) bands. We obtain converged results with respect to the number of Floquet Fourier basis using $n_{F}=150$ in the computations.  Figure \ref{fig:plot2}(a)-(c) show the optical absorption spectrum of the \emph{t}PA obtained using the absorption coefficient formula in Ref. \cite{Tiwari2023} with each absorption line broadened using a Lorentzian function of width 0.04 eV. The light-matter interactions are captured as in Eq. \eqref{nonlocalham} for both probe and drive laser. The different lines in each plot represent different number of bands taken (from 10 to 140 bands) into account in the computation. The inset in each plot shows the absorption spectrum in the low-frequency range ($\hbar \omega \in [0,0.5]$ eV) of the probe laser.  As seen, the absorption spectrum for $E_{\mathrm{d}}= 0.1 $ V/\r{A} require 30 bands (orange line) for convergence for $\hbar \omega \in [1,6]$ eV and 120 bands (purple line) in the low-frequency range.  Upon increasing the drive field amplitude to $E_{\mathrm{d}}= 0.2 $ V/\r{A},  70 bands (green line) are required for fully converged absorption spectrum in the [1,6] eV of the electromagnetic spectrum. Convergence in the low-frequency region still requires 120 bands. Further increasing the electric field amplitude to $E_{\mathrm{d}}= 0.3 $ V/\r{A} shows that even 100 bands are not enough for a converged calculation. Convergence in this case is achieved for 120 bands. Overall, the computations in Fig. \ref{fig:plot2} underscore the limitations of the velocity gauge consideration in Floquet engineering where a large number of bands are required for convergence. This requirement becomes increasingly more problematic as the drive field strength is increased.

We now show how our strategy to take into account space truncation solves these issues. We use as a benchmark, the 140 band computations as it yields converged results for the range of the drive field strength considered.

\subsection{Optical absorption spectrum of laser-dressed materials in truncated velocity gauge}\label{sec:resultstruncvelocity}

\begin{figure}[htbp!]
    \centering
    \includegraphics[width=0.48\textwidth]{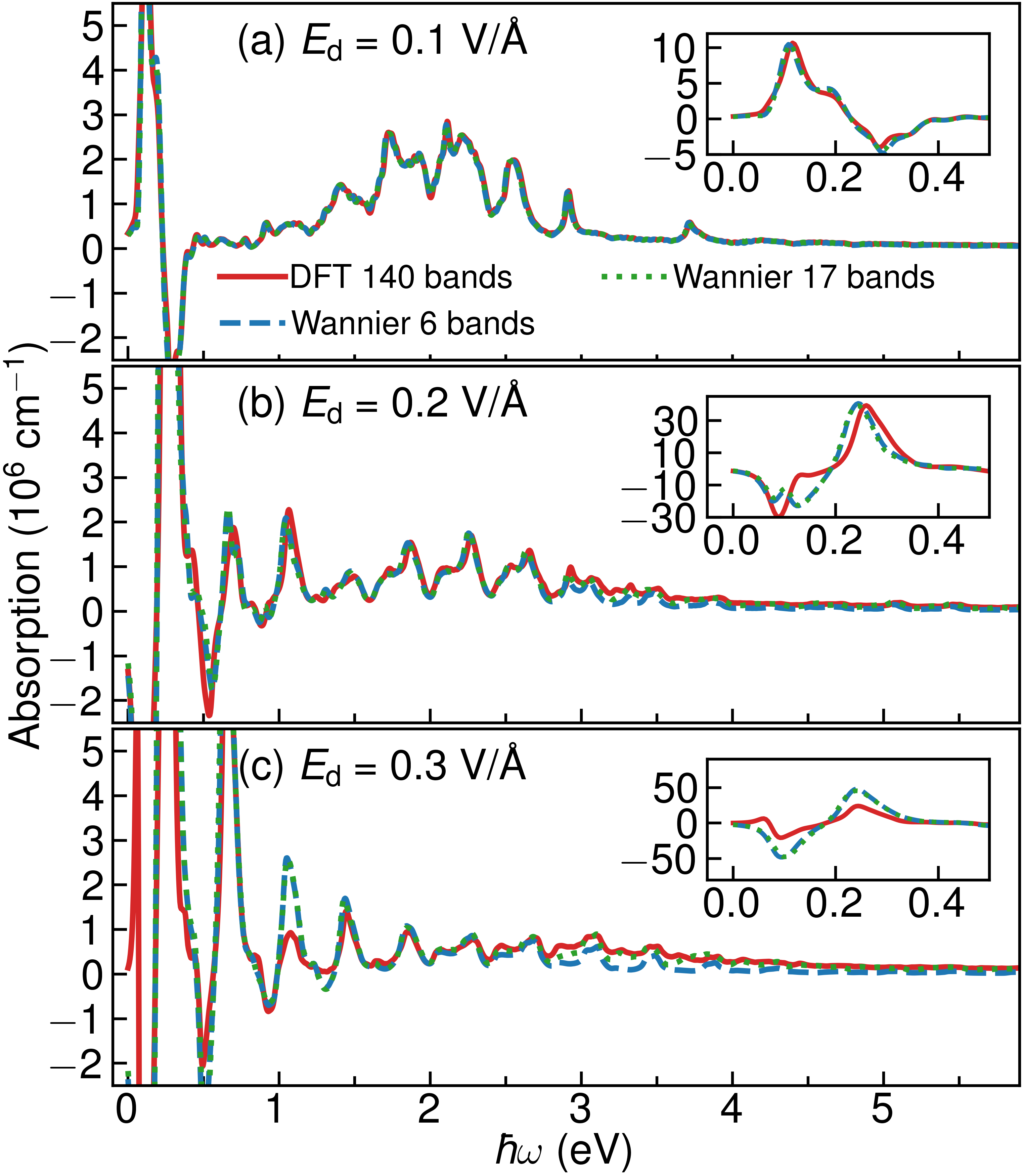}
    \caption{Truncated velocity gauge computations of the optical absorption spectrum of \emph{t}PA  for $E_{\mathrm{d}}$ (a) 0.1 V/\r{A} (b) 0.2 V/\r{A} and (c) 0.3 V/\r{A}, using a  6-band (blue lines) and 17-band (green lines) generalized tight-binding model. For comparison, computations in the velocity gauge with the 140-band DFT model (red lines) are also shown. Insets details the spectrum in the low-frequency region.}
    \label{fig:plot3}
\end{figure}

We now discuss the optical absorption spectra of the laser-dressed \emph{t}PA computed in the truncated velocity gauge as discussed in Sec. \ref{sec:theory}. The tight-binding parameters are obtained from Wannier interpolation for the 6- and 17-band model.  The first 3 bands (or first 5) in the 6 (or 17) Wannier band model correspond to valence band states. We use  $500 \times 1 \times 1$ $\mathbf{k}$-vector grid for the calculations, $n_{F}=170$, and up to 22 nested commutators in the laser-dressed Hamiltonian and in the truncated momentum matrix elements. The large number of commutators are required to capture the light-matter interactions due to the non-resonant driving pulse, as the importance of the matrix elements scale with the powers of $E_{\mathrm{d}}/\hbar\Omega$  and $\hbar\Omega=0.4$ eV is low in this case.

Figure \ref{fig:plot3}(a)-(c) shows the absorption spectra of the laser-dressed \emph{t}PA in the truncated velocity gauge using the 6- (blue lines) and 17-band (green lines) models for $E_{\mathrm{d}}=0.1-0.3$ V/\r{A}. These computations are contrasted with the optical absorption spectrum obtained from velocity gauge calculations with the 140-band DFT  model (red lines). The truncated velocity gauge computations significantly show faster convergence with the number of bands compared to the velocity gauge computations in Fig. \ref{fig:plot2}. For example, as seen in Fig. \ref{fig:plot3}(a) for $E_{\mathrm{d}} = 0.1$ V/\r{A}, the truncated velocity gauge calculation with 6-band is already converged in contrast to the 30 bands needed in the velocity gauge (Fig. \ref{fig:plot2}(a), orange line). The 6-band and 17-band truncated computation are also aligned on top of each other for $E_{\mathrm{d}} = 0.2$ V/\r{A} in Fig. \ref{fig:plot3}(b) while velocity gauge requires 70 bands for convergence (Fig. \ref{fig:plot2}(b), green line). Even at strong drive laser amplitude $E_{\mathrm{d}} = 0.3$ V/\r{A}, the 6-band spectrum  is converged in $\hbar \omega \in [0, 3]$ eV range while  velocity gauge computations in $\hbar \omega \in [0.5, 3]$ eV required 100 bands for convergence (Fig. \ref{fig:plot2}(c), brown line). Further, the 6-band computations are converged in the low-frequency region ($\hbar\omega \in [0, 0.5]$ eV) for the range of $E_{\mathrm{d}}$ studied here in Fig. \ref{fig:plot3}(a)-(c) where 120 bands were required in the velocity gauge computation (see Fig. \ref{fig:plot2}(a)-(c) purple line) for convergence. The deviations between the 6- and the 17-band model seen in Fig. \ref{fig:plot3}(b)-(c) in the $\hbar\omega > 3$ eV range are attributed to the missing bands in the computation.

Figure \ref{fig:plot3} also shows the accuracy of the truncated velocity gauge computations. As seen in Fig. \ref{fig:plot3}(a) for $E_{\mathrm{d}}=0.1$ V/\r{A}, the 6-band computation in the truncated velocity gauge reproduce the 140-band DFT model spectrum. Fig. \ref{fig:plot3}(b) for $E_{\mathrm{d}}=0.2$ V/\r{A}, the 140-band DFT model spectrum in velocity gauge shows some deviations with respect to the 6- and 17-band spectrum in truncated velocity gauge. We attribute these differences to the neglected terms in Eq. \eqref{nonlocalham} in the usual DFT approach that arise due to the non-local pseudopotential present in the realistic material Hamiltonian. The computations in the truncated velocity gauge for the 6- and 17-band model are more accurate in this case as no such approximation is involved. For strong drive  amplitude $E_{\mathrm{d}}= 0.3$ V/\r{A} in Fig. \ref{fig:plot3}(c), the DFT based computation in velocity gauge show larger deviations with respect to the truncated velocity gauge computations as expected because the neglected terms are non-linear in the drive.

The computations of the optical absorption spectrum in the truncated velocity gauge accurately recover the results of the velocity gauge using just a few bands. The truncated velocity gauge also shows faster converge  with respect to the number of bands compared to the velocity gauge. The velocity gauge computations work accurately for relatively low electric fields amplitude up to 0.2 V/\r{A} but additional contributions from the neglected terms in Eq. \eqref{nonlocalham} are required to capture the exact laser-dressed dynamics at higher field strengths. The converged calculations in the truncated velocity gauge are based on exact light-matter interaction Hamiltonian and are expected to accurately represent the theoretical absorption spectrum of the laser-dressed \emph{t}PA. 

In terms of the computational cost, the converged absorption spectrum of the laser-dressed \textit{t}PA for the truncated velocity gauge (17-band) takes 4 minutes of CPU time for calculations of the absorption lines for one $\mathbf{k}$-vector on the Intel Xeon Gold 6330 processor. In turn, the converged computations in the velocity gauge (140-band) take 1080 minutes. This shows that the truncated velocity gauge calculations are significantly faster than the velocity gauge calculations as needed to characterize the properties of laser-dressed materials using realistic models.

\section{Conclusions \label{sec:conclusion}}

To summarize, we developed a general strategy to capture Floquet engineering in solids in the velocity gauge using realistic Hamiltonians. Floquet considerations in the velocity gauge require a very large number of bands for convergence. Further, when the Hamiltonian for the realistic solid is constructed from first-principles, the non-local pseudopotential introduces non-linear light-matter coupling terms in the velocity gauge that are challenging to evaluate. As shown, these issues inherent to the velocity gauge make the Floquet engineering computations for realistic solids challenging.

We showed that it is possible to overcome these issues by explicitly taking into account the truncation of the Hilbert space in the light-matter interactions and by describing the material in terms of a generalized tight-binding model in the formulation of the theory for Floquet engineering. Hilbert space truncation replaces the momentum operator by a truncated momentum operator defined in terms of the nested commutator of position and material Hamiltonian. We exemplified the approach in the challenging case where there is both a drive and probe laser at play. Specifically, we applied this methodology to our theory of optical absorption of laser-dressed solids and developed analytical expression for its  optical absorption coefficient. The final formula is reminiscent to the optical absorption of pristine matter but with the Floquet-Bloch modes playing the role of pristine eigenstates of solid. Our results clarify how to effectively capture Hilbert space truncation in Floquet engineering in the velocity gauge.

To illustrate the methodology, we performed computations of the optical absorption spectrum of laser-dressed \emph{trans}-polyacetylene with both the usual velocity gauge and the truncated velocity gauge introduced here. While the velocity gauge required $140$ bands for convergence, the truncated velocity gauge yielded accurate results with just 6 bands. For strong driving amplitudes, in practice, velocity gauge  computations can only be considered approximate due to the large number of bands needed and the additional non-linear terms in light-matter interactions introduced by the non-local pseudopotential. By contrast, the truncated velocity gauge calculations based on generalized tight-binding models, provide converged results in models of reduced dimensionality and fully capture the light-matter interactions. Overall, using our proposed strategy, we were able to satisfactorily integrate the Floquet engineering in velocity gauge with a realistic description of a solid. This work introduces an efficient theoretical tool to simulate Floquet engineering in realistic materials.


\begin{acknowledgments}
The authors thank Bing Gu and Sobhit Singh for helpful discussions. This material is based upon work supported by the National Science Foundation under Grant No. CHE-2416048.
\end{acknowledgments}

\appendix*

\begin{widetext}

\section{Matrix elements of the nested commutators}

The matrix elements to be evaluated in Eq. \eqref{commutmme} are of the form
\begin{align}
\label{evalone}
& \mu_{ff' \mathbf{k}} = \frac{1}{i} \langle \psi_{f \mathbf{k}} ^{(\mathrm{W})} |  [  \hat{\mathbf{r}} , \hat{\mathcal{O}}  ] | \psi_{f' \mathbf{k}} ^{(\mathrm{W})}\rangle ,
\end{align}
where $\hat{\mathcal{O}}$ is a space-periodic operator. The Bloch-like states can be expressed as a Fourier series in the MLWF basis as $|\psi_{f\mathbf{k}}^{(\mathrm{W})} \rangle = \frac{1}{\sqrt{N}}\sum_{\mathbf{R}} e^{ i \mathbf{k} \cdot \mathbf{R}} |f \mathbf{R} \rangle$ which yields
\begin{align}
\label{evaltwo}
& \mu_{ff' \mathbf{k}} = \frac{1}{N} \sum_{\mathbf{R},\mathbf{R}'} \frac{1}{i} e^{i\mathbf{k} \cdot (\mathbf{R'}-\mathbf{R})} \bigl( \sum_{f'' , \mathbf{R''}} \langle f \mathbf{R} |   \hat{\mathbf{r}} |  f'' \mathbf{R''}  \rangle \langle  f'' \mathbf{R''} | \hat{\mathcal{O}}  |   f' \mathbf{R'}  \rangle   - \langle f \mathbf{R} | \hat{\mathcal{O}} |  f'' \mathbf{R''}  \rangle \langle  f'' \mathbf{R''} | \hat{\mathbf{r}}  |   f' \mathbf{R'} \rangle \bigr) ,   
\end{align}
where we have used  $\sum_{f''\mathbf{R} ''}|  f'' \mathbf{R''}  \rangle \langle  f'' \mathbf{R''} | = \hat{1}$ assuming a complete Wannier basis.  We consider the position operator matrix elements in the Wannier function basis given by  $ \langle  f \mathbf{R} | \hat{\mathbf{r}}  |   f' \mathbf{R'} \rangle  = \mathbf{R} \delta_{ff'} \delta_{\mathbf{RR'}} + \langle  f \mathbf{0} | \hat{\mathbf{r}}  |   f' \mathbf{R'} -\mathbf{R} \rangle  $ \cite{Pedersen2001}. Substituting these matrix elements into Eq. \eqref{evaltwo} yields
\begin{align}
    \nonumber
     \mu_{ff' \mathbf{k}} & =  \frac{1}{N} \sum_{\mathbf{R},\mathbf{R}'} (-i) e^{i\mathbf{k} \cdot (\mathbf{R'}-\mathbf{R})} \biggl[ \sum_{f'' , \mathbf{R''}}  \bigl( \mathbf{R} \delta_{\mathbf{R,R''}} 
 \delta_{f,f''} + \langle  f \mathbf{0} | \hat{\mathbf{r}}  |   f'' \mathbf{R''} -\mathbf{R} \rangle \bigr) \langle  f'' \mathbf{R''} | \hat{\mathcal{O}}  |   f' \mathbf{R'} \rangle  \biggr. \\ 
 \nonumber
  & \biggl. - \langle f \mathbf{R} | \hat{\mathcal{O}} |  f'' \mathbf{R''}  \rangle \bigl( \mathbf{R}' \delta_{\mathbf{R'',R'}} 
 \delta_{f'',f'} + \langle f'' \mathbf{0} | \hat{\mathbf{r}}  |  f' \mathbf{R' - R''} \rangle \bigr)  \biggr]  \\
 \nonumber
 & = \frac{1}{N} \sum_{\mathbf{R},\mathbf{R}'} i (\mathbf{R}'- \mathbf{R} ) e^{i\mathbf{k} \cdot (\mathbf{R'}-\mathbf{R})}  \langle  f \mathbf{R} | \hat{\mathcal{O}}  |   f' \mathbf{R'} \rangle    + \frac{1}{N} \sum_{\mathbf{R},\mathbf{R}'} (-i) e^{i\mathbf{k} \cdot (\mathbf{R'}-\mathbf{R})}  \\ 
 & \bigl( \sum_{f'' , \mathbf{R''}}  \langle  f \mathbf{0} | \hat{\mathbf{r}}  |   f'' \mathbf{R''} -\mathbf{R} \rangle  \langle  f'' \mathbf{R''} | \hat{\mathcal{O}}  |   f' \mathbf{R'} \rangle - \langle f \mathbf{R} | \hat{\mathcal{O}} |  f'' \mathbf{R''}  \rangle \langle f'' \mathbf{0} | \hat{\mathbf{r}}  |  f' \mathbf{R' - R''} \rangle \bigr)  .
 \end{align}
In the above equation, we substitute $i(\mathbf{R}-\mathbf{R}') e^{i \mathbf{k \cdot (R-R')}} = \nabla_{\mathbf{k}} e^{ i \mathbf{k \cdot (R-R')}}$ and define the matrix elements of the periodic operator as  $\mathcal{O}_{ff'}^{(\mathrm{W})}(\mathbf{k}) = \frac{1}{N} \sum_{\mathbf{R}} e^{i\mathbf{k} \cdot \mathbf{R}} \langle f \mathbf{0} | \hat{\mathcal{O}}| f' \mathbf{R} \rangle =  \frac{1}{N} \sum_{\mathbf{R,R'}} e^{i\mathbf{k} \cdot (\mathbf{R'-R}) } \langle f \mathbf{R} | \hat{\mathcal{O}} | f' \mathbf{R'} \rangle $ to get
 \begin{align}
 \nonumber
 &  \mu_{ff' \mathbf{k}} =  \nabla_{\mathbf{k}}  \mathcal{O}_{f f'}^{(\mathrm{W})}(\mathbf{k})  \\ 
 \nonumber
 & - i \sum_{\mathbf{R},\mathbf{R}' , f'' , \mathbf{R''}}    \bigl(   \langle  f \mathbf{0} | \hat{\mathbf{r}}  |   f'' \mathbf{R''}  - \mathbf{R} \rangle e^{i\mathbf{k} \cdot (\mathbf{R'' - R})} \mathcal{O}_{f''f'}^{(\mathrm{W})}(\mathbf{k})  - \mathcal{O}_{ff''} ^{(\mathrm{W})}(\mathbf{k}) \langle f'' \mathbf{0} | \hat{\mathbf{r}}  |  f' \mathbf{R' - R''} \rangle e^{i\mathbf{k} \cdot (\mathbf{R'} -\mathbf{R''})} \bigr) \\
 &  = \nabla_{\mathbf{k}}  \mathcal{O}_{f f'}^{(\mathrm{W})}(\mathbf{k})  -i \bigl( \sum_{ f'' } A_{ff''} ( \mathbf{k}) \mathcal{O}_{f''f'}^{(\mathrm{W})}(\mathbf{k}) -  \mathcal{O}_{ff''}^{(\mathrm{W})} ( \mathbf{k}) A_{f'' f'}(\mathbf{k}) \bigr) =  \nabla_{\mathbf{k}}  \mathcal{O}_{f f'}^{(\mathrm{W})}(\mathbf{k})  -i   [ \hat{A} , \mathcal{\hat{O}}  ]_{ff'\mathbf{k}} 
  \end{align}
  which is right hand side of Eq. \eqref{commutmme}, and where  $A_{ ff''}( \mathbf{k}) \equiv \sum_{\mathbf{R}} e^{i\mathbf{k} \cdot \mathbf{R}} \langle f \mathbf{0} | \hat{\mathbf{r}} | f'' \mathbf{R} \rangle $ is the Fourier-transformed position operator.

\end{widetext}

\bibliography{gaugeinv}

\end{document}